\documentclass[journal=jmcmar,layout=twocolumn]{achemso}

\author{Michal H. Kol\'{a}\v{r}}
\affiliation[IOCB CAS CR]
{Institute of Organic Chemistry and Biochemistry of the Czech
Academy of Sciences, Flemingovo nam. 2, 16610 Prague, Czech
Republic}
\email{michal@mhko.science}
\altaffiliation{Current address: Department of Theoretical and
Computational Biophysics, Max Planck Institute for Biophysical
Chemistry, Am Fassberg 11, D-37077 G\"ottingen, Germany}
\author{Oriana Tabarrini}
\affiliation{Department of Pharmaceutical Sciences, Universita
of Perugia, Via del Liceo 1, I-06123 Perugia, Italy}

\title{Halogen Bonding in Nucleic Acid Complexes}

\abbreviations{IR,NMR,UV}
\keywords{DNA, RNA, ribosome, $\sigma$-hole, sigma-hole, 
electrostatic potential, Protein Data Bank, drug design}

\begin{document}



\begin{abstract}
Halogen bonding (X-bonding) has attracted notable attention among
noncovalent interactions. This highly directional attraction 
between a halogen atom and an electron donor has been exploited
in knowledge-based drug design. A great deal of information has been
gathered about X-bonds in protein-ligand complexes, as opposed to
nucleic acid complexes. Here we provide a thorough analysis
of nucleic acid complexes containing either halogenated building
blocks or halogenated ligands. We analyzed close contacts between halogens
and electron-rich moieties. The phosphate backbone oxygen is clearly 
the most common halogen acceptor. We identified 21 X-bonds within known
structures of nucleic-acid complexes. 
A vast majority of the X-bonds is formed by halogenated nucleobases,
such as bromouridine, and feature excellent geometries.
Noncovalent ligands have been found to form only
interactions with sub-optimal interaction geometries. Hence,
the first X-bonded nucleic-acid binder remains to be discovered.
\end{abstract}


\section{Introduction}

Bringing a new drug to the market consumes enormous intellectual and
financial resources. It typically takes more than a decade from the
discovery of an active compound (lead) to the final approval of 
a drug, which is based on it. The drug discovery and development
declined from the trial-and-error approach when the number of
recognized diseases steeply rose. Nowadays, the workflow stems
from knowledge gained in a variety of studies believing that this
can speed-up the whole drug development.

The number of known drug targets is slightly higher than 
300 \cite{Overington06}.
This amount is not extraordinarily high bearing in mind the number
of recognized human genes (less than 20,000 \cite{Ezkurdia14}). 
Among the targets, there are 
mostly proteins and only a few nucleic acids (NAs).
Given that NAs are ubiquitous biopolymers with a myriad 
of cellular functions though, they represent a clinically prominent 
class of targets \cite{Overington06}.

Many strategies have appeared to optimize the lead compound into
a therapeutic substance. The use of halogens is in this sense
traditional. Xu \emph{et al}. estimated that
about 25 \% of approved drugs are halogenated, and the portion
is similar in all stages of drug discovery and
development \cite{Xu14}.
Halogen atoms modulate physicochemical properties
of the molecular scaffolds; they affect the polarity and
hydro/lipophilicity, which in turn changes membrane and 
blood brain barrier permeation of the molecule. Also, 
carbon-halogen covalent bond is difficult to metabolize,
so the halogenation prolongs the lifetime of the active
compound, but at the same time might increase its liver 
toxicity \cite{Hernandes10}.

Apart from nonspecific effects, it was about a decade ago 
recognized that halogens might partake in a structurally 
specific and directional noncovalent interaction called 
a halogen bond (X-bond)\cite{Cavallo16}.
The X-bond is an interaction between a halogen and a Lewis base or an
electron-rich moiety. The electron density donors may be represented 
by electronegative atoms such as oxygen, nitrogen, sulfur, but also 
by aromatic rings or conjugated $\pi$-systems.

The X-bond has been found in many protein-ligand complexes including
pharmaceutically relevant ones (for review see Refs. \citenum{Ford15},
\citenum{Scholfield13}, and \citenum{Wilcken13}). There have been 
only a few studies on X-bonds, where NAs played a role. The
distinguished exceptions are the efforts of Shing Ho and co-workers.
They focused on halogen bonding in so-called Holliday junction,
a four-stranded (branched) complex of deoxyribonucleic acid (DNA)
\cite{Duckett88, Lilley00}. Using bromouridine as a building block,
they directed a DNA into one of the several nearly
isoenergetic conformers \cite{Voth07}. The DNA model system also
demonstrated that among the halogens it is the bromine that has the
most favorable entropy/enthalpy compensation. Consequently, bromine
was claimed an optimal element for X-bonding in DNA Holliday
junctions \cite{Carter11, Carter13}.

To the best of our knowledge, no ligand has been reported so far
to form an X-bond in any NA complex.
The lack of information on
X-bonding in NAs is somewhat surprising because NAs are naturally
rich in electronegative atoms which make them (in theory) 
prospective X-bond acceptors. It seems that there is
a missing relation between the worlds of NAs and X-bonds. This work
aims at building such a relation. Hence, we continue with
the introductions of the two worlds trying to find some overlap 
between them and highlight the pharmaceutical significance. Later, 
we analyze known structures of NAs and reveal main features of 
their complexes. Finally, we discuss all of the few examples
of low-molecular compounds whose halogens are involved in interactions
with electron donors.

\section{Halogen Bonding Features}

\begin{figure}
\includegraphics{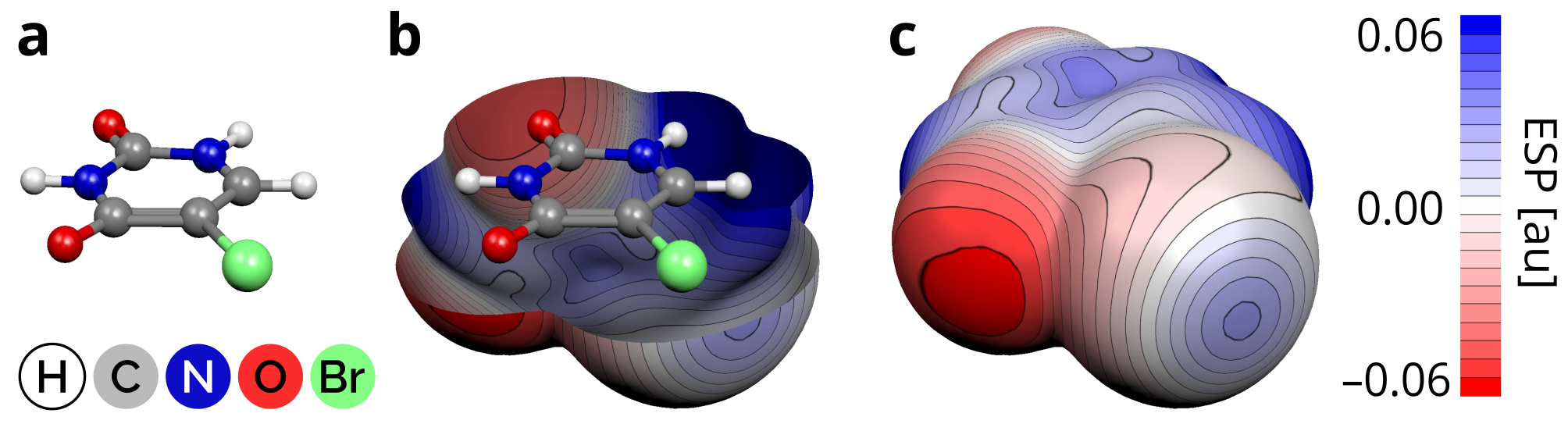}
\caption{Molecular structure of bromouracil (a), the intersection 
of a molecular surface (b), and full ESP in atomic units (au)
projected onto the molecular surface (c).
The blue disc of positive ESP in the forefront is the $\sigma$-hole.}
\label{fig:sigmahole}
\end{figure}

The attraction of halogens with other electronegative atoms was
observed as early as the 1950s in the crystallographic studies of
Hassel \emph{et al.} \cite{Hassel54} although the synthesis
of X-bonded complexes dates back to 19th century \cite{Guthrie63}. 
The puzzling nature of the attraction between two electron-rich
chemical groups was attributed merely to charge transfer effects
until Politzer \emph{et al.} came up with a simple model explaining
many of the X-bond features \cite{Brinck92}. Based on quantum 
chemical calculations of the molecular electrostatic potentials
(MEPs) they proposed that the surface of halogens contains 
regions of both positive as well as negative electrostatic 
potential (ESP) \cite{Brinck92}. The positive region was labeled
a sigma-hole ($\sigma$-hole) (Fig. \ref{fig:sigmahole}),
and interestingly enough this label 
appeared only 15 years after its first evidence \cite{Clark07}. 
The X-bond has been exploited in many areas of chemistry and
material science (reviewed e.g. in Refs. \citenum{Cavallo16} and
\citenum{Gilday15})
and a great amount of work has been done on the theoretical aspects
of X-bonds too \cite{Kolar16, Ford15}.

X-bonds are similar in strength to the more common hydrogen bonds
(H-bonds). The X-bond stabilization energy typically amounts 
to 5--25 kJ/mol \cite{Rezac12, Kolar14a} and it increases with the 
increasing atomic number of the halogen involved \cite{Politzer10}.
The reason for this is the increasing size and magnitude of the
halogen $\sigma$-hole \cite{Kolar14b}. Another factor is the halogen 
polarizability \cite{Riley08}, which also increases with the halogen
atomic number. Contrary to traditional view, modern theoretical
studies suggest that the role of charge transfer in the X-bond
stabilization is modest \cite{Rezac17}. The strongest X-bonds
are found in complexes of iodinated molecules; brominated and chlorinated
molecules form weaker X-bonds. Fluorine is the least polarizable and the 
most electronegative halogen. It possesses positive $\sigma$-holes in rare
cases and mostly in inorganic molecules \cite{Metrangolo11}, 
so it is of lower importance for biological applications.

The structural trends of X-bonds in biomolecules have been inferred
from several Protein Data Bank (PDB) \cite{Berman00} surveys and
theoretical analyses \cite{Auffinger04, Lu10, Scholfield13,
Sirimulla13}. All of the studies focused on protein-ligand X-bonds.
The X-bonds are mostly established with the protein backbone. Its
carbonyl oxygen is the most frequent electron donor \cite{Lu10}. 
No preference for backbones of $\alpha$-helices, $\beta$-sheets 
or loop structures has been found \cite{Wilcken12}. 
Lu \emph{et al.} reported that about one-third of protein-ligand
X-bonds involve an aromatic ring of the amino acid side chains
\cite{Lu10}. It was also reported that the X-bonds might not 
disrupt existing networks of H-bonds \cite{Rowe17}. Instead, an orthogonal 
pattern of X- and H-bonds is preferred involving the same electron
donor atom \cite{Voth09}. Another effect of ligand halogenation is
a shortening of proximal H-bonds \cite{Poznanski14}.

\section{Nucleic Acids as Drug Targets}

The deoxyribonucleic and ribonucleic acids (DNA and RNA) appear in
living organisms in several forms. Whereas the DNA adopts only a few
conformational classes, the structural diversity of RNA is much
broader. DNA mostly occurs in the B-form helical conformation. A key
RNA motif is the A-form double helix. However, loop regions, bulges,
and other forms of mismatched nucleobases often disrupt the motif. 
Such a conformational variability is mirrored in the rapidly 
expanding variety of RNA functions recognized, especially in 
the last two decades. Apart from the classic roles of ribosomal,
transfer and messenger RNAs, RNA was also shown to store genetic
information, regulate gene expression, or act as 
enzyme \cite{Kiss02, Mercer09}.

Many NA-binders are halogenated. The DNA alkylating agents often
contain halogen atoms because halogens facilitate
the alkylating reactions on NAs. This way, anticancer cisplatin
and its analogues \cite{Rosenberg65, Weiss93} covalently modify 
DNA.
The reaction products block replication or transcription processes.
The effect of alkylation is non-specific in terms of DNA sequence
and also cell type. The modifications occur in both normal and 
cancer cells, but the higher proliferation rate of the cancer 
cells makes these drugs effective.

Alternatively to covalent modifications, various classes of compounds
bind to double-stranded DNA (dsDNA) in a noncovalent manner. Examples
include the anticancer anthracycline type antibiotics daunomycin and
adriamycin, and the polypeptide antibiotic dactynomicin, which works
mainly by intercalating into DNA. The intercalation, described as an
insertion of a planar molecule between consecutive base pairs,
interferes with DNA-processing enzymes \cite{Baguley91}. Other small
molecules bind to dsDNA interacting with base pair functional groups
on the floor of the minor or major groove. Much of the research has
concentrated on the DNA minor groove recognition leading to improved
sequence selectivity of the compounds \cite{Baraldi04, Barrett13}.

Important classes of drugs target ribosomal RNA (rRNA). In fact, 
most of the RNA-targeting drugs on the market act on
ribosomes \cite{Thomas08}. The ribosome is a biomachine which
synthesizes proteins. As such, it represents a critical center
of cellular life. The differences between bacterial and eukaryotic
ribosomes have allowed developing specific antibacterial agents 
that are often based on natural products \cite{Hermann05,
Poehlsgaard05, Wilson14}. There is a variety of binding locations
within the ribosome; the most frequent sites are 
the peptidyl-transferase center on the 50S subunit and the decoding
center on the 30S subunit. Aminoglycosides and tetracyclines are 
the best-known classes of small molecules whose primary target 
is the 30S subunit. Oxazolidonones instead exert their 
antibacterial action by binding to the 23S rRNA present within 
the 50S subunit. Macrolides and related compounds (lincosamides
and streptogramins), as well as chloramphenicol and clindamycin
also target the 50S subunit. Several ribosome-binding molecules 
have been prepared containing a halogen atom stemming mostly from
the pioneering case study of chloramphenicol \cite{Ehrlich47}.

In the past decade, other non-coding RNAs (ncRNAs) have emerged
as prospective drug targets \cite{Thomas08, Hermann16}.
Highly conserved ncRNAs provide new opportunities to expand the
repertoire of drug targets to treat infections. Viral regulatory
elements located in untranslated regions of mRNA often form folded
structures that harbor potential binding sites for small 
molecules. The absence of homologous host cell RNAs makes them
attractive for the development of innovative antiviral compounds.

There are a plethora of ncRNAs under intense pharmaceutical 
research due to their ability to interact with low-molecular 
ligands. For instance, human immunodeficiency virus type-1 (HIV-1)
Trans-activation response (TAR) RNA plays an essential role in HIV-1
replication through its interaction with the viral trans-activator of
transcription. Such interaction might be disrupted by ligands 
(reviewed in Refs. \citenum{Mousseau15} and \citenum{Tabarrini16}),
where some of them contain halogen atoms \cite{Mayer04}.

Another example is an internal ribosome entry site (IRES) from
Hepatitis C virus (HCV) \cite{Lukavsky09}. The IRES RNA contains
several independently folding domains that are potential targets 
for the development of selective viral translation inhibitors. 
A diverse set of ligands including oligonucleotides, peptides
as well as small molecules have been reported to block IRES 
function by distinct mechanisms \cite{Dibrov13}. Like in the case 
of other ncRNAs, the drug candidates occasionally contain halogens.
Nevertheless, no drug targeting a ncRNA has been approved for the
market.

Many lines of evidence are linking mutations and dysregulations of
ncRNAs to neurodegenerative disorders \cite{Castel10, Mirkin07}.
The presence of expanded CNG repeats in 5' and 3' untranslated 
regions is related to important diseases such as Myotonic dystrophy
type 1, spinocerebellar ataxia type 3, and fragile X-associated 
tremor-ataxia syndrome. For example, the pathological expansion 
of CAG repeats ($>$35 consecutive CAG codons) in huntingtin 
exon 1 encodes a mutant protein whose abnormal function 
determines Huntington's disease. Finding compounds able 
to bind pathogenic CNG repeats with high specificity may be
a valuable strategy against these devastating diseases. Their
design is still limited by the lack of structural information,
although some small molecules have emerged through various 
strategies \cite{Kumar12, Bochicchio15}.

Both DNA and RNA can also adopt a non-canonical higher-order 
structure called G-quadruplexes (G4s) that are involved in 
regulating multiple biological pathways such as transcription,
replication, translation and telomere structure \cite{Lipps09}.
The building blocks of G4s are guanosine-rich quartets that
self-associate into a square-planar platform through a cyclic
Hoogsteen H-bonded arrangement. G4s are found in oncogene 
promoters, in telomeres, as well as in introns of mRNAs. 
These regions have been recognized as potential targets 
for anticancer drugs \cite{Han00}. A large number of small
molecules are able to bind the quadruplex structures. They
are characterized by polycyclic heteroaromatic scaffolds, 
or by cyclic/acyclic non-fused aromatic rings. Thus far, only
a few molecules have been found to selectively bind the 
telomeric G4s, although their therapeutic potential appears
high \cite{Neidle09}.

\section{Nucleic Acids as X-bond Acceptors}

In the X-bond, the $\sigma$-hole on a halogen represents a Lewis 
acid that interacts with a Lewis base. NAs seem to offer an 
abundance of basic chemical groups. The backbone contains phosphate groups
with two oxygens carrying a charge of $-1$. Likewise, ribose and
deoxyribose contain oxygens with lone electron pairs that could
serve as $\sigma$-hole acceptors too. Further, the nucleobases
form H-bonds that could be potentially replaced or augmented by
X-bonds.

\begin{figure*}
\includegraphics{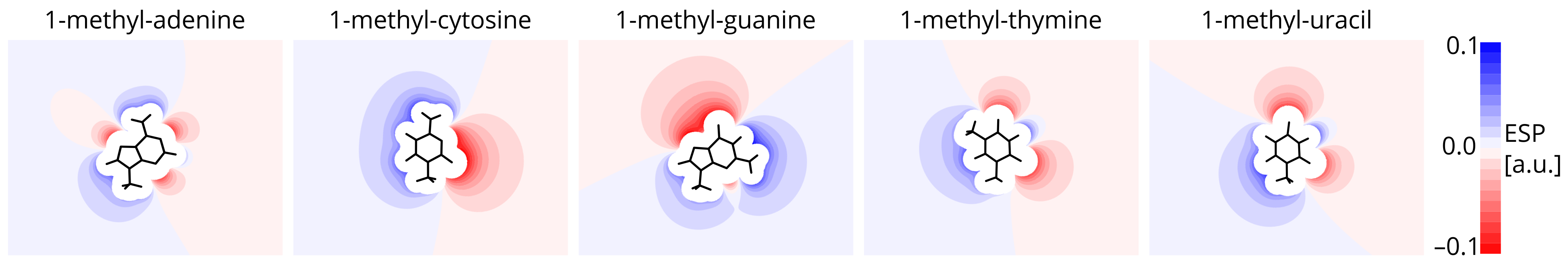}
\caption{2D projections of the electrostatic potential (ESP)
of the most common nucleobases A, C, G, T, U onto the planes
of their aromatic rings. Note the ESP in atomic units (au)
was calculated for the N-methylated analogs.}
\label{fig:espbases}
\end{figure*}

Nucleobases themselves show certain electrostatic diversity, where 
the negative sites may play a role of halogen acceptors.
Fig. \ref{fig:espbases} depicts MEPs of five most common nucleobases
projected onto the plane of their aromatic systems. Cytosine and
guanine contain areas of more negative ESP than the other 
nucleobases. Thymine and uracil resemble each other having two
negative sites on the oxygens separated by a small positive site.
Perhaps the most heterogeneous ESP is around adenine which exhibits
six areas of zero ESP near the molecular surface, as compared to 
four (G, T, U) and two (C). Apart from this, all of the nucleobases
are aromatic, which allows them to act as electron donors via their
$\pi$-electrons above and below their rings (not shown).

In nucleic acids, the situation is more complicated than in 
nucleobases: the three-dimensional structure of DNA and RNA is
electrostatically diverse due to the presence of the phosphate
backbone. For instance, the DNA minor groove was shown to be more
electronegative than the major groove \cite{Chin99}. Thanks to its
plasticity, RNA may create folds with even more unusual 
electrostatic characteristics. Indeed, regions of strong negative
electrostatic potentials were exploited in designing efficient TAR
binders \cite{Davis04}. Electrostatic interaction is also the main
driving force of aminoglycoside binding to the ribosome \cite{Ma02}.
Overall, NAs seem to offer favorable electrostatics to attract 
positive $\sigma$-holes on halogens. It remains elusive, how is 
such ability employed in ligand recognition and NA self-assembly.

\section{Structural Survey Yields Two Data Sets}

To understand interaction preferences of halogens
we analyzed known NA structures. We started with 
a broad set of X-ray structures from
the PDB (September 2016) \cite{Berman00}. Apart from
X-ray structures, the PDB also contains NA structures determined by
nuclear magnetic resonance (NMR) techniques. For two reasons, we 
deliberately omitted those from our analyses. First, we wanted 
to be consistent with the strategies adopted by previous structural
surveys of X-bonds in protein-ligand complexes.
Second, there are indications 
that it may be difficult to assess the quality of the deposited NMR
structural ensembles \cite{Nabuurs06}, which could complicate the
geometric characterization.

Within the X-ray structures, we searched for complexes containing
a nucleic acid and a halogen atom (Cl, Br, or I). Fluorine was 
excluded from the search due to its extremely low ability to form
X-bonds in biological systems. The selected structures comprise 
nucleic acids and their complexes with other nucleic acids,
low-molecular ligands, and/or proteins. To get reliable geometric
characteristics, only data with the resolution better than 
3.0 \AA~were considered. Note that the previous PDB surveys
of X-bonds in protein complexes used the same resolution threshold
\cite{Auffinger04, Sirimulla13}.

From the PDB, we obtained 672 files which were subsequently filtered.
We excluded structures containing halogens in the form of ions.
Following the recommendation of the International Union of Pure 
and Applied Chemistry \cite{Desiraju13}, we selected
X-bonds as the contacts between halogens and electronegative
atoms (N, O, P) shorter than or equal to the sum of van
der Waals (vdW) radii \cite{Bondi64} (Tab. \ref{tab:radii}).
The interactions involving at least one of the two interacting
atoms with the crystal occupancy lower than 0.5 were omitted. 
Because of the X-bond directionality\cite{Kolar14b}, 
we also required the angle of 
R--X$\cdots$Y to be higher than 120$^\circ$. Same or similar
geometric criteria were used previously to define biological
X-bonds with proteins \cite{Auffinger04, Lu10}, although 
the wider angular range was used in other studies as well
\cite{Scholfield13, Sirimulla13}.

\begin{table}
\begin{tabular}{llll}
\hline
\hline
    & N    & O    & P \\
\hline
Cl  & 3.30 & 3.27 & 3.55 \\
Br  & 3.40 & 3.37 & 3.65 \\
I   & 3.53 & 3.50 & 3.78 \\
\hline
\hline
\end{tabular}
\caption{Sums of van der Waals radii \cite{Bondi64} of atomic pairs
in \AA.}
\label{tab:radii}
\end{table}

We also searched for the X-bonds that are formed with 
the aromatic systems of the nucleobases. To this aim, 
we considered only halogen contacts closer than 5 \AA~to
the aromatic plane that make an angle between the plane 
normal vector and the X--C bond smaller than 60$^\circ$.


In the end, we obtained a set of 21 X-bonds that satisfy the data quality,
chemical and geometric criteria. This amount is a rather low.
Scholfield \emph{et al.} reported 760 protein complexes with an X-bond
in 2012, which stands for about 1 \% of the 80,000 protein structures
deposited in the PDB at that time. The 21 X-bonds here represent about
0.2 \% of ca 9,000 structures containing NA.

Hence to better capture possible geometric properties of halogen
interactions, we collected a more extended set of complexes with
a longer interaction distance. An arbitrary threshold of 4 \AA~was
chosen such that it is higher than the X-bond length but still
short enough to hint for an attractive interaction.
Within the article, the interactions are referred 
to as \emph{linear contacts} and they comprehend the X-bonds 
as well. We found 72 linear contacts.
Table \ref{tab:pdbcounts} summarizes various subsets of the PDB query.

\begin{table*}
\begin{tabular}{lrrrr}
\hline
\hline
 & Cl & Br & I & sum \\
\hline
Files & & & & \\
PDB search count &
402 & 204 &  66 &  672 \\
Contains contact(s) &
43  & 162 &  34 &  239 \\ 
Contains contact(s) with NA building block &
31  & 135 &  22 &  188 \\  
Contains linear contacts(s) with NA building block &
22  & 29  & 2  & 53 \\ 
Contains X-bond(s) with NA building block &
2   & 17  & 2  & 21 \\ 
\hline
Interactions & & & & \\
Number of contacts &
611    & 1,319 & 205  & 2,135 \\ 
Contacts with NA building block &
43     & 315  & 41   & 399 \\ 
Linear contacts with NA building block &
22     & 48   & 2    & 72 \\ 
X-bonds with NA building block &
2      & 17   & 2    & 21 \\ 
\hline
\hline
\end{tabular}
\caption{Analysis of the PDB. The contact is defined by the
interatomic distance between a halogen and oxygen, nitrogen,
or phosphorus shorter than 4 \AA. The linear contact has
the angle R--X$\cdots$O/P/N higher than 120$^\circ$. X-bonds
is a linear contact shorter than the sum of vdW radii.}
\label{tab:pdbcounts}
\end{table*}

\section{The X-Bonds Favor Nucleo-base$\cdots$Phosphate Pattern}

Within the X-bond set, the variety of interacting partners is low;
20 X-bonds involve a halogenated nucleobase,
one X-bond is formed by cisplatin chlorine. The set contains
19 X-bonds with a phosphate oxygen as the electron donor;
further, there is one X-bonds with an aromatic ring, and one with
cytosine oxygen. Overall, only two X-bonds do not concur
with the dominant nucleobase$\cdots$phosphate interaction pattern.

The X-bond geometries are close to ideal. The X-bond lengths
are shorter than the sum of the vdW radii by about 8 \%, which
conforms with the contractions reported for the protein X-bonds
\cite{Scholfield13}. About 90 \% of the X-bonds are 
straighter than 160$^\circ$.

\begin{figure}
\includegraphics{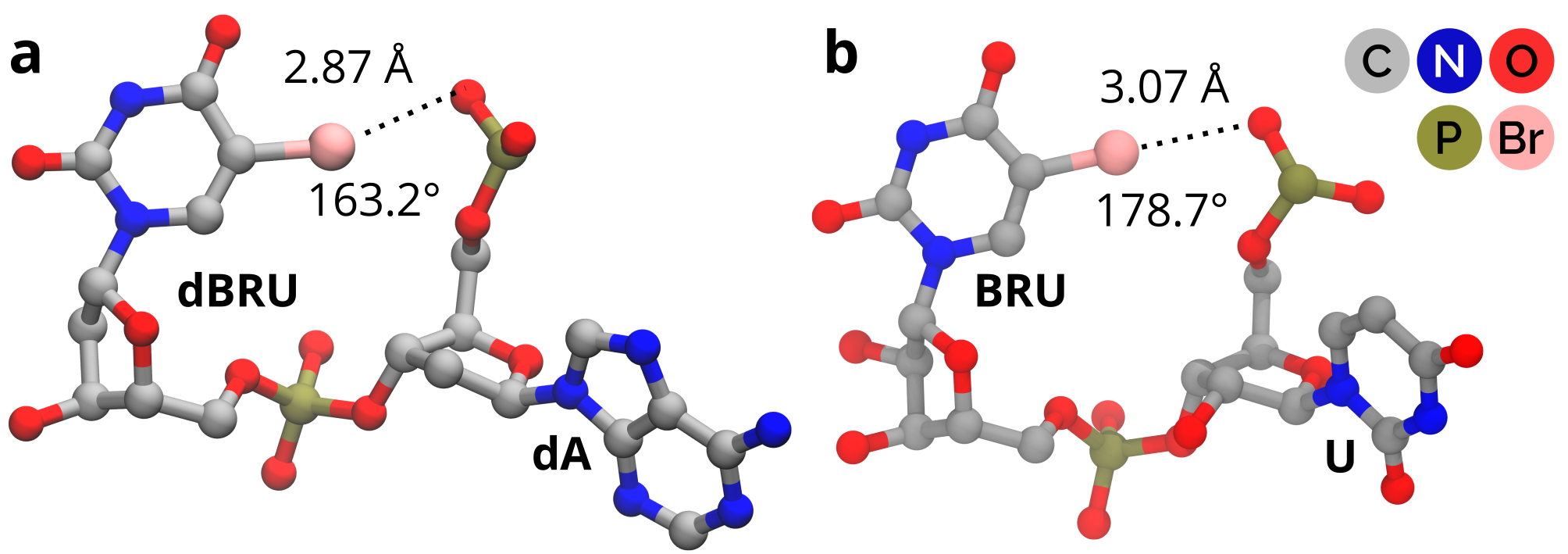}
\caption{The shortest (a) and the 
straightest (b) X-bonds found. Hydrogens were
omitted for 
clarity. Residues are labeled as BRU: 5-bromouridine, dBRU: 
5-bromo-2’-deoxyuridine, U: uridine, dA: deoxyadenosine.}
\label{fig:records}
\end{figure}

The shortest X-bond in the set belongs to a structure of
a Holliday junction; the X-bond is found between 
a bromodeoxyuridine and a phosphate oxygen of two neighboring 
residues (Fig. \ref{fig:records}a) (PDB: 2org, resolution
2.0~\AA) \cite{Voth07}. The X-bond was shown to stabilize 
particular DNA assembly in competition with H-bond. 
The straightest X-bond also involves a brominated uridine
and phosphate oxygen (Fig. \ref{fig:records}b) (PDB: 2bu1, 
resolution 2.2~\AA) in a complex of RNA and phage MS2 coat 
protein \cite{Grahn01} and the overall geometry is remarkably
similar to the one found in Holliday junction. Whereas the 
study on Holliday junction fully appreciated the role of
X-bonding, in the latter case the specific interaction of
the bromine remained unrecognized.

\begin{figure}
\includegraphics{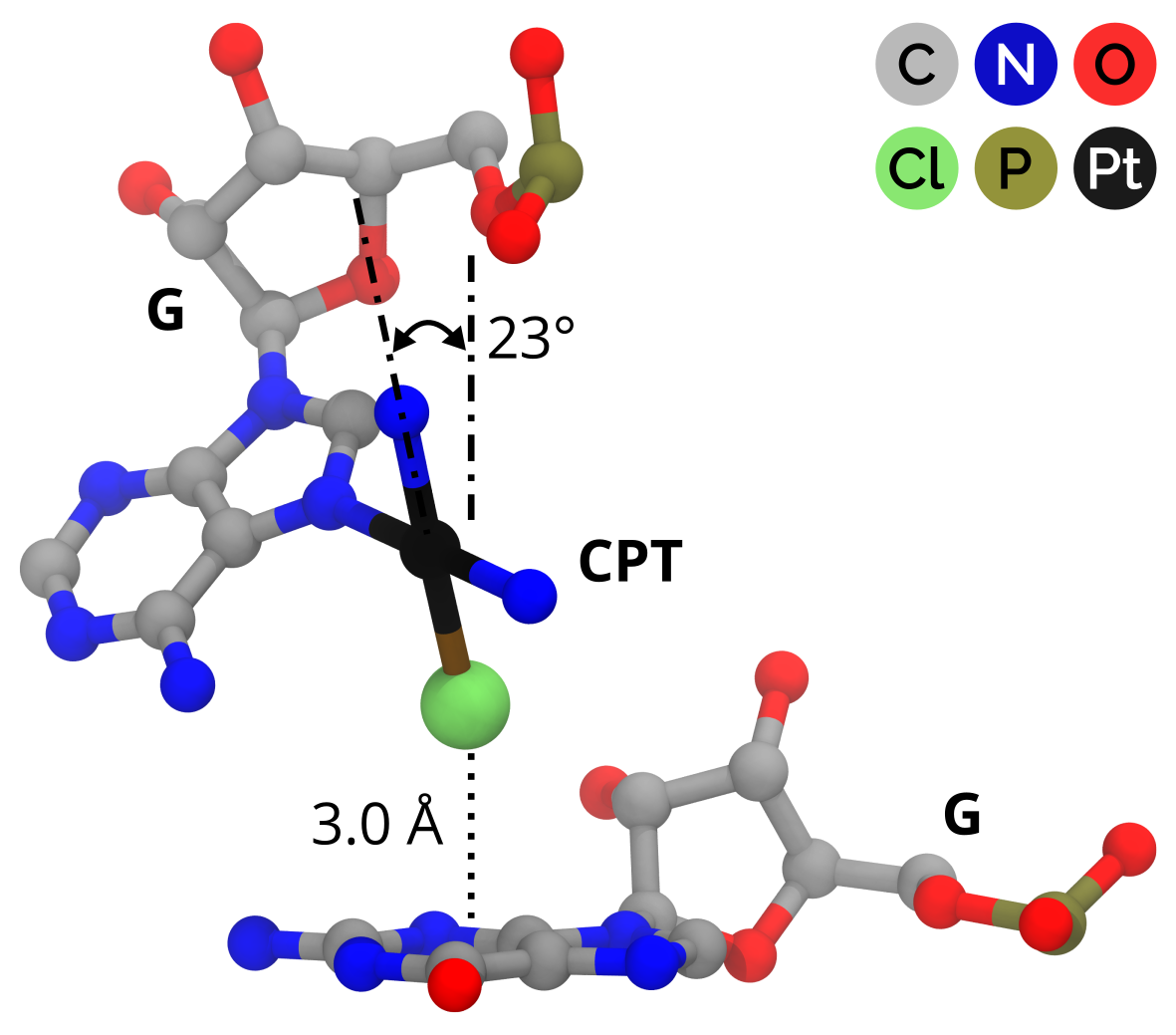}
\caption{X-bond with an aromatic system between 
cisplatin (CPT) and a guanine (G). The distance stands for 
the perpendicular distance from the aromatic plane.}
\label{fig:aromatic}
\end{figure}

What is important, we have identified no noncovalent ligand involved
in X-bonding.
The only non-nucleobase residue that participates in an X-bonds
is cisplatin covalently bound to an adenine (PDB 5j4c,
resolution 2.8 \AA \cite{Melnikov16}).
It is also the only halogen donor that forms an X-bond with 
a $\pi$-system, although the interaction with one of the guanine nitrogens
would alone classify the interaction as X-bond.
The X-bond features superb geometric characteristics 
(Fig. \ref{fig:aromatic}). Unlike proteins, in nucleic acids, the
halogen-$\pi$ interaction competes with $\pi-\pi$ interactions
more often. Especially in dsDNA, it is hardly conceivable that 
there is a space for a halogen to attack the nucleobases from 
above or below of their aromatic planes. The situation in RNAs
might be more favorable, but the single occurrence of such X-bond
is hard to generalize.

\section{Interactions Longer Than X-Bonds}

Table \ref{tab:charac} summarizes counts of various types of interactions,
and the statistics of the interaction geometries of the set of 72 halogen
linear contacts.

In the NA complexes, the geometric quality of the halogen interactions
increases in the order of Cl $<$ Br $<$ I. The interaction angles are 
more linear for heavier halogens. Especially the contacts of chlorine
are rather bent with the median angle of 141$^\circ$. The medians of
the interaction lengths (Tab. \ref{tab:charac}) decrease with
the increasing atomic number of the halogen. The surveys on 
protein-ligand X-bonds \cite{Auffinger04, Lu10} revealed the
opposite trend, i.e. the increasing length of X-bonds with 
the increasing atomic
number of the halogen involved. In this work (but also in Ref.
\citenum{Auffinger04}), the statistical sample might be
insufficient to provide reliable statistics, which is true 
especially for the two iodine X-bonds.

\begin{table*}
\begin{tabular}{lrrr}
\hline
\hline
Halogen & Count & Length [\AA] (med$\pm$iqr) & Angle [deg] (med$\pm$iqr) \\
\hline
chlorine               & 22 & 3.49$\pm$0.31 & 141$\pm$8 \\
bromine                & 48 & 3.45$\pm$0.30 & 165$\pm$22 \\
iodine                 & 2  & 3.07$\pm$0.06 & 173$\pm$3 \\
\hline
Electron Donor & & &  \\
\hline
N                      & 21 & 3.47$\pm$0.28 & 140$\pm$11 \\
non-backbone O         &  9 & 3.69$\pm$0.42 & 135$\pm$25 \\
backbone O             & 42 & 3.41$\pm$0.45 & 166$\pm$8 \\
\hline
Halogenated Residue & & & \\
\hline
Halogenated nucleotide & 51 & 3.43$\pm$0.43 & 165$\pm$23 \\
Low-molecular ligand   & 21 & 3.50$\pm$0.29 & 140$\pm$7 \\
\hline
\hline
\end{tabular}
\caption{Medians (med) and interquartile ranges (iqr) 
of geometric characteristics of interactions involving various
types of partners.}
\label{tab:charac}
\end{table*}

\begin{figure}
\includegraphics{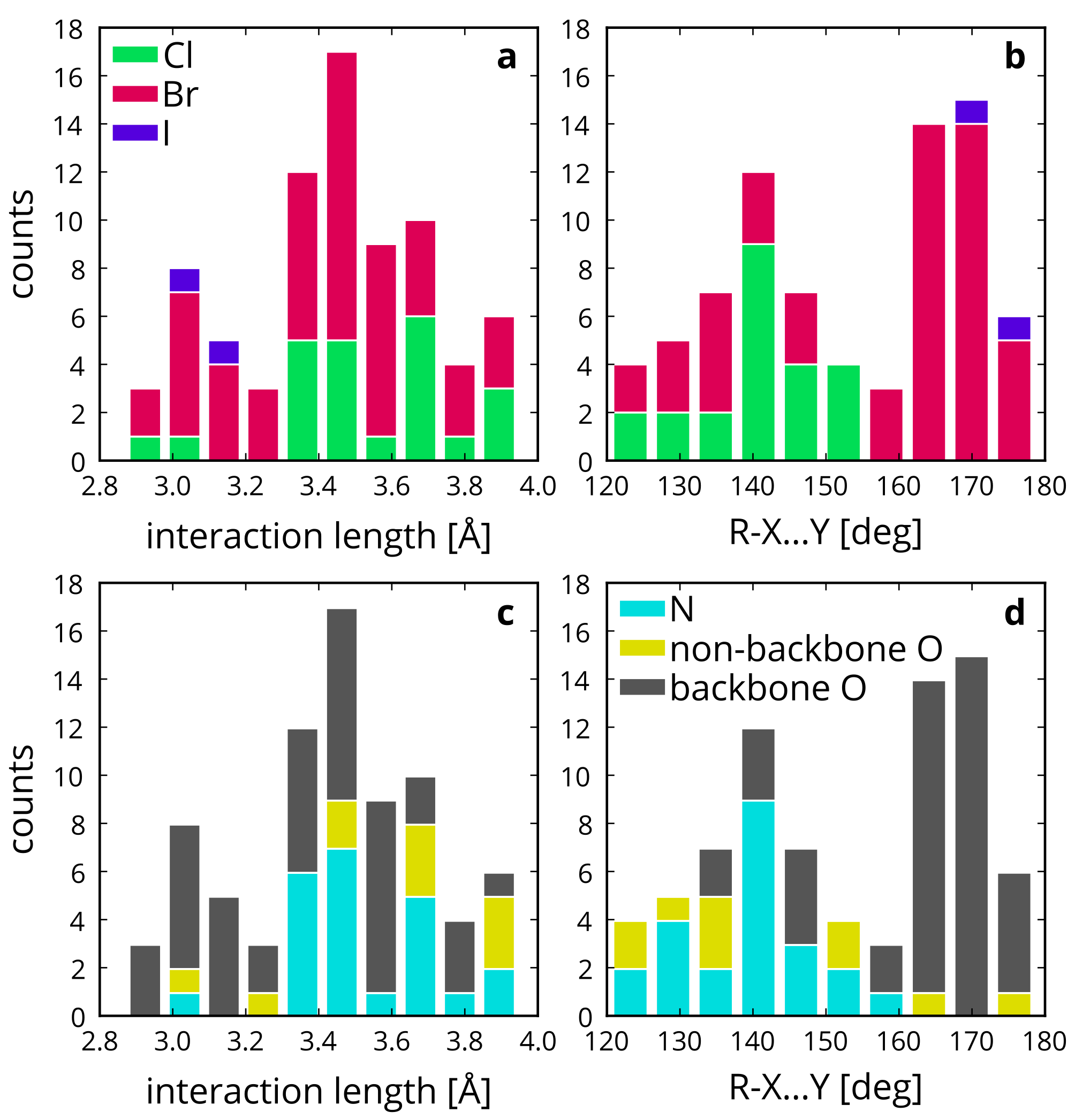}
\caption{Histograms of interaction lengths (a, c) and interaction angles 
represented by the R--X$\cdots$Y angle (b, d); color-coded according to the
halogen involved (a, b), or electron donating atom (c, d).}
\label{fig:characteristics}
\end{figure}

Fig. \ref{fig:characteristics} shows the histograms of the interaction
lengths and angles. In the histogram of lengths, the minor peak 
near 3.0~\AA, which is comparable with the sum of the vdW radii,
stands for almost ideal X-bond length. There is a major peak
near 3.5~\AA~too. Two peaks also appear in the histogram of interaction 
angles; around 140$^\circ$ and 170$^\circ$.

\begin{figure}
\includegraphics{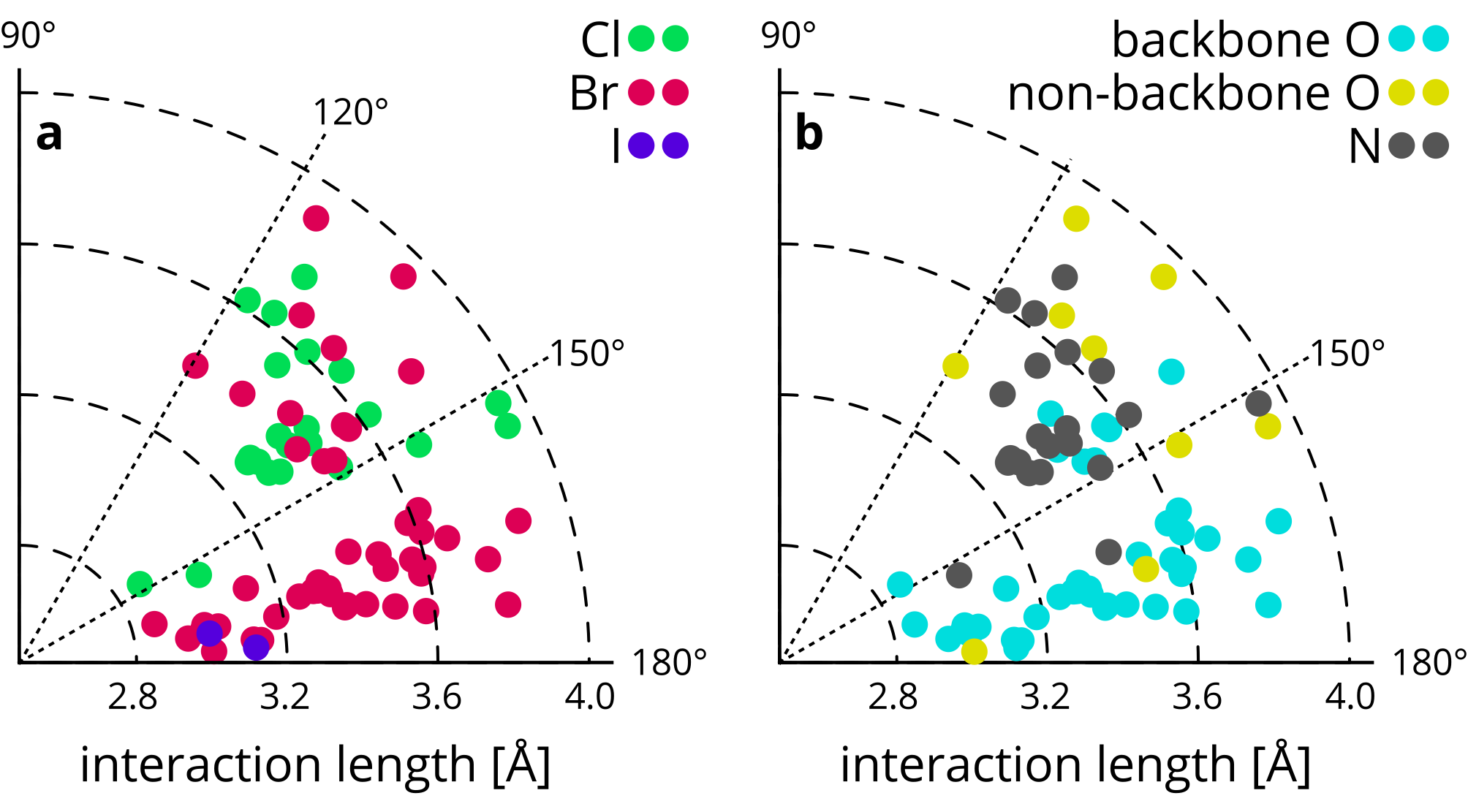}
\caption{Interaction geometries in polar coordinates color coded
according to the halogen (a), or the electron donor atom (b).}
\label{fig:polar}
\end{figure}

Each of the halogen subsets contributes to the histograms with
a different weight. 
Chlorine- and bromine-containing complexes 
span the whole range of interaction lengths. The two
iodine complexes feature short interactions. The situation with 
angles is different. Chlorine complexes appear only in the region
of lower interaction angles. The highest angle in the chlorine subset is
154$^\circ$, so chlorine interactions are likely weak. 
On the other hand, bromine complexes are
scattered across the whole range of angles (Fig.
\ref{fig:characteristics}b) with a cumulation
around 170$^\circ$ (Fig. \ref{fig:polar}).
The two iodine X-bonds are very straight (X-bond
angle higher than 170$^\circ$) suggesting a strong interaction.

We analyzed the electron donors of halogen interactions
found in the NA complexes. The backbone oxygen atom is the most
common one. We identified 
51 linear contacts of halogens with oxygen (71~\%), and 
21 with nitrogen (29~\%). Although we included 
phosphorus as an electron donor in the search, no linear contact
was found. The actual occurrence of the oxygen and
nitrogen in NAs is roughly 2:1 (O:N), which likely
contributes to the dominance of the linear contacts with oxygen. Most of
the nitrogen interactions were found with chlorine, whereas bromine
preferentially interacts with oxygen. 82~\% of the all
interacting oxygens belong to the phosphate backbone. Unlike
other oxygens and nitrogens in NAs, the phosphate oxygens carry
a negative charge, which explains their higher propensity to
halogen $\sigma$-holes.

According to the electron-donating atoms, the geometric
quality of the interactions increases in the order of
non-backbone oxygen $<$ nitrogen $<$ backbone oxygen 
(Tab. \ref{tab:charac}). 
The interactions which employ a backbone oxygen are typically shorter
and straighter than the others.

We conclude that different interaction atoms 
likely occur in different interaction geometries.
It is also apparent on a projection of the interaction
geometries to the polar coordinates (Fig. \ref{fig:polar}).
We observed no correlation between interaction lengths and the
corresponding angles.

\section{Halogenated Ligands Show Sub-Optimal Interaction Geometries}

We did not find any X-bonded noncovalent ligands in NA complexes.
Nevertheless, in the set of linear contacts we have found 15 unique 
ligands -- 14 chlorinated and one brominated. The remaining interactions 
involve halogenated nucleobases as halogen donors.
The halogen interactions of ligands are longer and less linear compared
to the interactions of halogenated nucleobases (Tab. \ref{tab:charac}).
The interaction angles deviate notably from linearity, 
which suggest that such interactions do not play a critical role
in the NA-ligand recognition.

\begin{figure}
\includegraphics{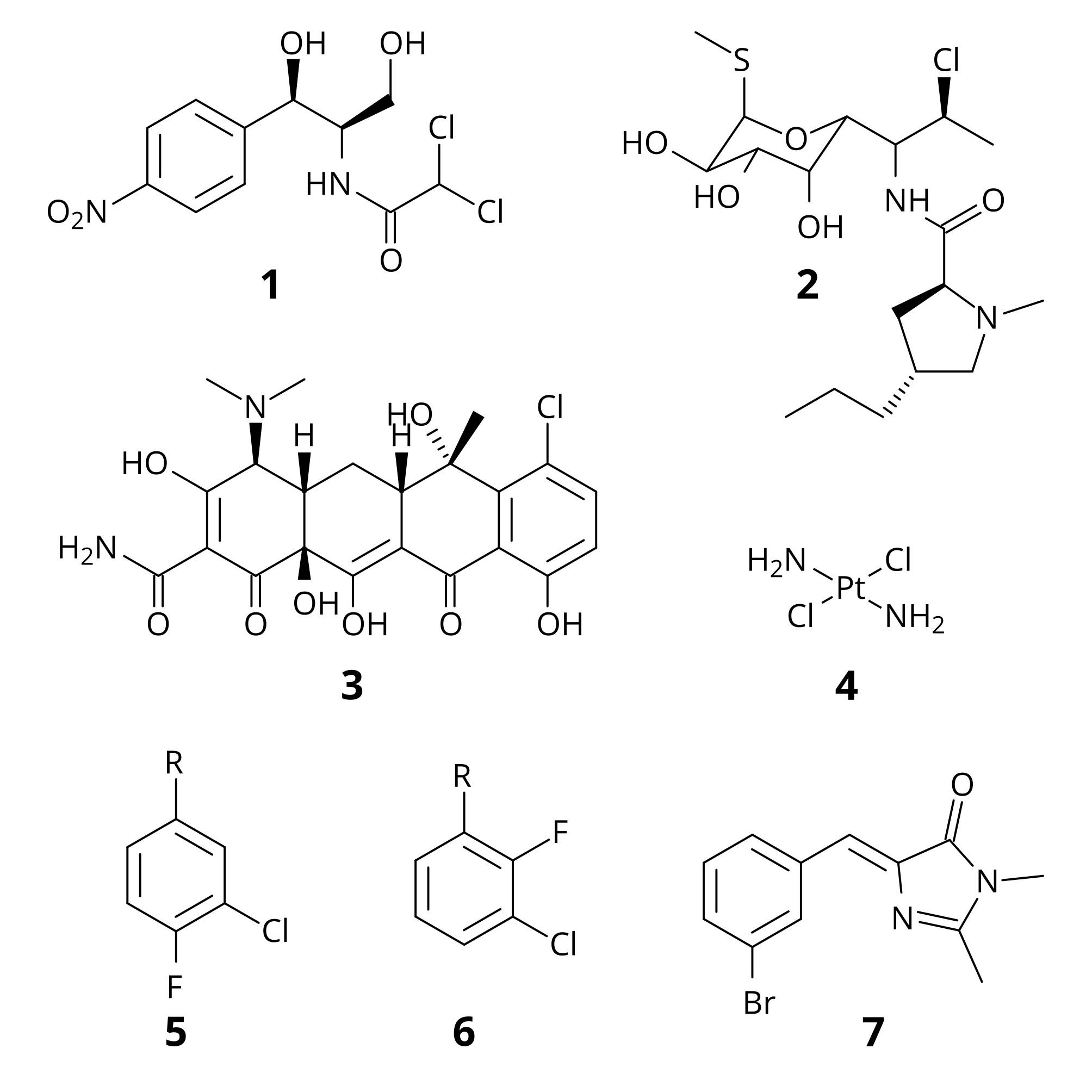}
\caption{Structural formulas of several low-molecular 
ligands or their building blocks that participate in halogen 
interactions.}
\label{fig:formulas}
\end{figure}

From the pharmaceutical point of view, the low-molecular ligands
are of higher interest than the halogenated building blocks. Below
we discuss all of the instances among the linear-contact
data set which involve a low-molecular ligand.
Structural formulas of the ligands 
discussed are shown in Fig. \ref{fig:formulas}.

\begin{figure}
\includegraphics{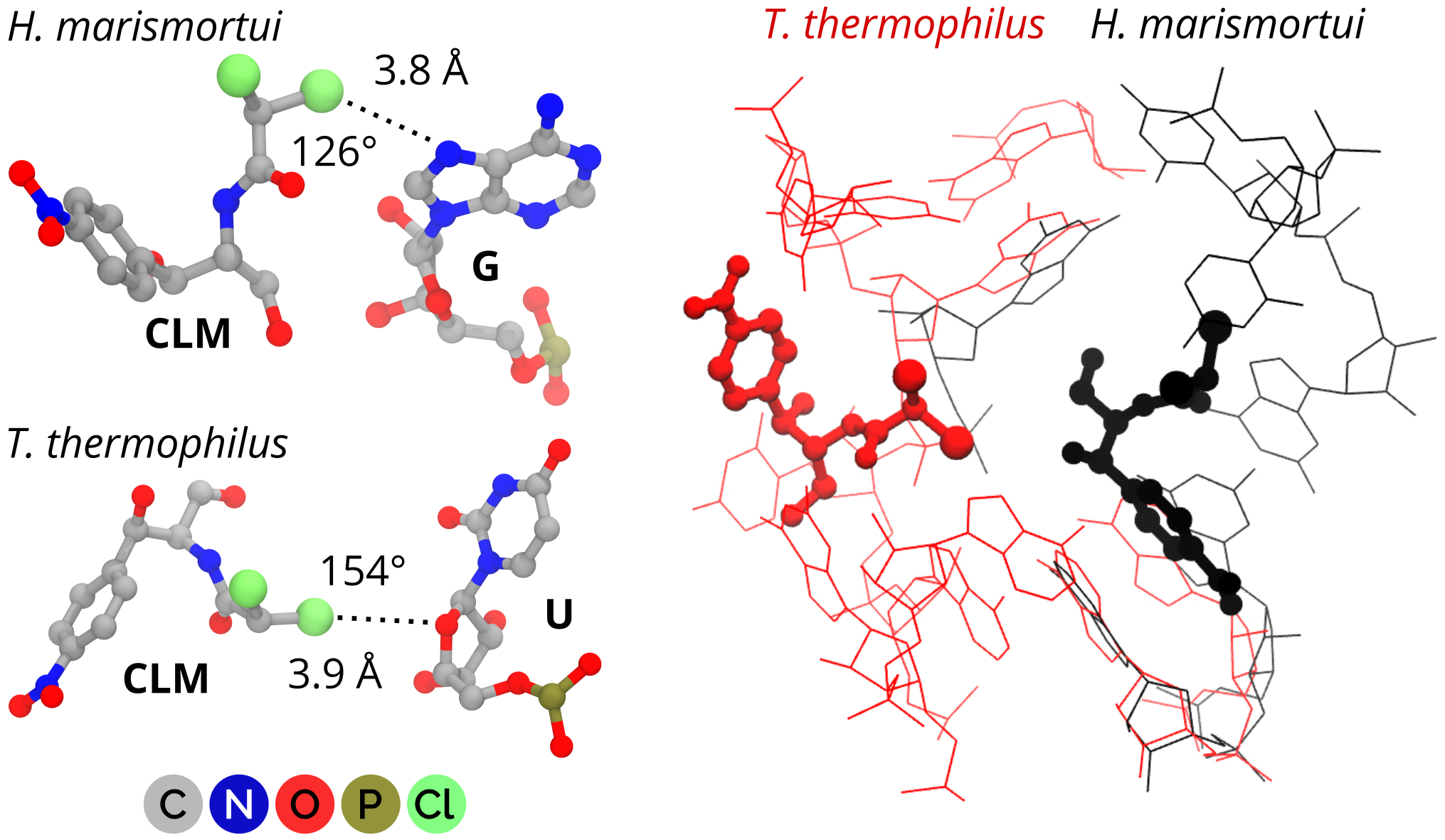}
\caption{Chloramphenicol \textbf{1} (CLM) interactions in the 50S ribosomal 
subunit with guanosine (G) and uridine (U). Hydrogens were
omitted for clarity. The overlay of the two binding poses found
in \emph{T. thermophilus} (PDB 1nji) and \emph{H. marismortui}
(PDB 4v7w) into a single reference frame is shown in red and 
black, respectively. Only residues within 7 \AA~from each of 
the halogens are shown.}
\label{fig:clm}
\end{figure}

Several halogenated compounds bind to the bacterial
ribosome. Chloramphenicol (\textbf{1}) \cite{Ehrlich47}
is a classic antibiotic
compound that targets the ribosomal A-site crevice in the 50S
subunit (PDB  1nji, resolution 3.0~\AA \cite{Hansen03}; 4v7w,
resolution 3.0~\AA \cite{Bulkley10}). \textbf{1} contains two
aliphatic chlorines that are activated by the nearby carbonyl
group. Two distinct binding orientations in two different ribosomal
system were proposed (\emph{T. thermophilus} \cite{Hansen03} 
and \emph{H. marismortui} \cite{Bulkley10}) (Fig. \ref{fig:clm}).
The lengths of both chlorine interactions are
beyond the respective sums of the vdW radii (about 3.3 \AA,
Tab. \ref{tab:radii}), 
and quite bent. One of the chlorines approaches either
the ribose in-ring oxygen (3.9 \AA, 154$^\circ$), or a guanine
nitrogen (3.8 \AA, 126$^\circ$), respectively. The geometries suggest
that the halogens interactions contribute weakly to the complex stability.

\begin{figure*}
\includegraphics{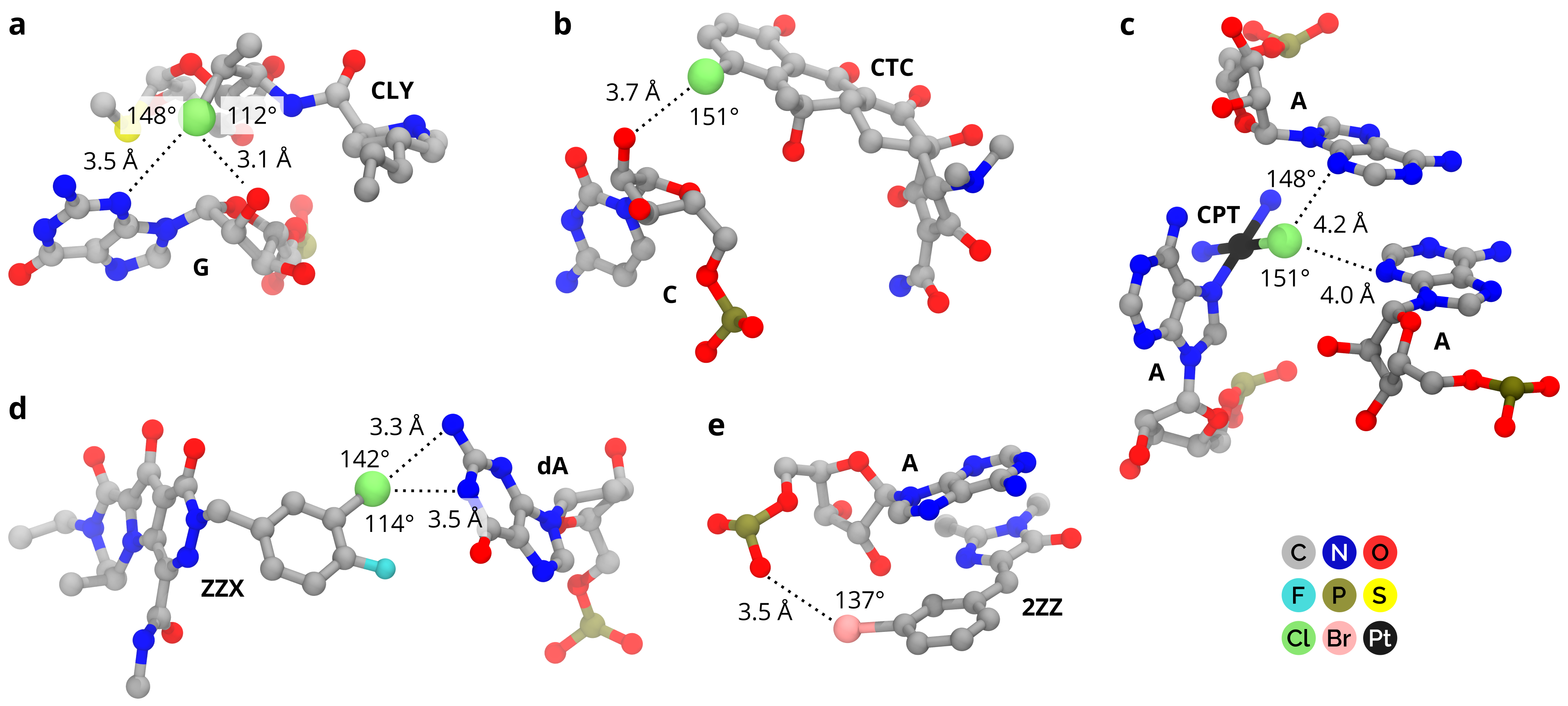}
\caption{Selected halogen interactions of ligand complexes.
Clindamycin \textbf{2} (CLY) bound to the ribosome (a),
chlorotetracycline \textbf{3} (CTC)
bound to an RNA aptamer (b), cisplatin \textbf{4} (CPT)
bound to ribosome (c), an intasome 
binder (ZZX) (d), a brominated imidazole analogue \textbf{7} (2ZZ) in 
an RNA aptamer (e). Hydrogens were omitted for clarity.
The interactions lengths and R--X$\cdots$Y angles are provided.}
\label{fig:xbs}
\end{figure*}

Clindamycin (\textbf{2}) \cite{Birkenmeyer70} is
another halogenated ribosome binder
\cite{Tu05} (PDB 1yjn, resolution 3.0~\AA) that binds into the
50S subunit (Fig. \ref{fig:xbs}a) and forms linear contacts.
\textbf{2} contains one chlorine that directs towards the sugar edge of
guanosine. There is an interaction with guanine nitrogen (3.5~\AA,
148$^\circ$). At the same time, there is a shorter but more 
bent contact with sugar O2' oxygen (3.1 \AA, 112$^\circ$).

7-chlorotetracycline (\textbf{3}) \cite{Duggar48} belongs to a class of 
tetracycline antibiotics that bind into the ribosomal 30S subunit.
This class acts by preventing correct processing of aminoacyl-tRNA.
Although no complex of \textbf{3} with the ribosome satisfies 
the data-set criteria, a complex with an RNA aptamer does. 
The aptamer was designed to bind \textbf{3} with sub-nanomolar
affinity (PDB 3egz, resolution 2.2~\AA \cite{Xiao08}). It 
features an interaction between a chlorine and a ribose oxygen 
of a cytidine (Fig. \ref{fig:xbs}b) (3.7~\AA, 151$^\circ$). Due to the 
sub-optimal geometry, the role of halogen interaction in complex
stabilization is likely marginal. Moreover, there are many other
intermolecular interactions between the \textbf{3} and the 
aptamer, such as a stacking of a planar part of \textbf{3} with
a guanine, and an oxygen-magnesium coordination (not shown). 
Those reduce the relative importance of the halogen contact even more.

Recently, crystal structures of \emph{T. thermophilus} ribosomes 
with cisplatin \cite{Rosenberg65} (\textbf{4} in 
Fig.~\ref{fig:xbs}c) were resolved
(PDB 5j4b, resolution 2.6~\AA; 5j4c, resolution 2.8~\AA \cite{Melnikov16}).
Nine molecules of cisplatin are covalently
bound to the ribosome in place of one of the chlorines. Two of the
cisplatin residues were identified to interfere with the mRNA
tunnel and the GTPase site -- the places critical for the ribosome
functions \cite{Melnikov16}. One cisplatin out of the nine forms an
X-bond with a guanine $\pi$-system (Fig.~\ref{fig:aromatic}).
Another cisplatin directs with its chlorine in between two stacked
adenines with the shortest distance to a nitrogen of 4.0~\AA ~and
angle 151$^\circ$. Another nitrogen is about 4.2~\AA ~far away and
forms the angle of 148$^\circ$ (Fig. \ref{fig:xbs}c).

Our data set contains several ligands that bind into the prototype
foamy virus (PFV) intasome, i.e. a complex of a viral integrase
and DNA. Such a system serves as a model for HIV-1 integrase
inhibition. Several inhibitors were developed with a halogenated
phenyl ring (\textbf{5} and \textbf{6}) that form X-bonds with
NAs \cite{Hare10, Metifiot12, Raheem15}. No difference in
the interaction geometries was found
between \textbf{5} and \textbf{6} analogs. \textbf{5}-analog
interactions show the best geometric characteristics among the PFV 
ligands (but generally still modest); the chlorine on
\textbf{5} interacts with two adenine nitrogens 
(Fig. \ref{fig:xbs}d). While the contact length of 3.3~\AA ~is
comparable with
the sum of vdW radii, the angle of 142$^\circ$ is far from the 
ideal 180$^\circ$. The other contact is 3.5~\AA-long and even more
bent (angle about 114$^\circ$).
We add that an aromatic fluorine activates the 
neighboring chlorine by making 
its $\sigma$-hole more positive \cite{Kolar14b}. 
The effect of fluorination on the ligand binding affinity is not
straightforward, however \cite{Fanfrlik13}.

We found only one brominated ligand with a contacts shorter than 4 \AA.
A brominated imidazole analog (\textbf{7}) binds into
a G-quadruplex formed within an RNA aptamer called \emph{Spinach} 
(PDB 4q9q, resolution 2.45~\AA) \cite{Huang14}. The Spinach 
module allows a green-fluorescence-protein-like functionality
by selectively activating fluorescence of \textbf{7}.
\textbf{7} binds in a planar conformation with a strong 
stacking interaction with adenine. Besides this, it forms a contact
with a phosphate oxygen of 3.5~\AA ~under the angle
of 137$^\circ$ (Fig. \ref{fig:xbs}e).

\section{Summary and Outlook}

The role of X-bonding in drug development has been emphasized
for a decade. Nonetheless, the knowledge-based design of 
halogenated drugs that feature an X-bond with their biomolecular
targets is still a tedious task. Although there have been published
several pioneering studies that involved protein targets
\cite{Llinas10, Hardegger11, Fanfrlik13}, the nucleic acids still 
lack a successful application. 

In this work, we focused on X-ray structures of biomolecular
complexes containing halogen interactions with nucleic acids.
The general ability of nucleic acids to 
provide electron donating groups to X-bonding has been confirmed here.
Using criteria recommended by IUPAC for X-bond definition, we found
21 NA complexes with the X-bond. We also analyzed interactions of 
the halogen atoms longer than X-bonds.

All but one of the X-bonds involved a halogenated nucleobase 
(preferably uridine). Halogenated nucleotides are often 
utilized in X-ray crystallography, which explains
their high occurrence in the NA complexes. The incorporation 
of heavy atoms such as bromine into the structure helps to 
overcome the phase problem. Moreover, the halogenated nucleotides
alter the physico-chemical properties of the NAs thus modulate 
the crystallization conditions; in the classical theory this 
is due to their stacking properties \cite{Bugg69}, but recently
X-bonding has also been shown to play a role \cite{Voth07}.
What is more, halogenated nucleobases may be also used in radiation 
anti-cancer therapy as radiosensitizers \cite{Erikson63}. Although,
the exact mechanism of the increased radiosensitivity is not known,
the changes in the ionization potential brought by the halogens
or global structural changes associated with modified nucleobase 
pairing have been proposed \cite{Wetmore01, Heshmati09}.

An important feature of X-bonds in NA complexes is their preference
towards backbone oxygens. This is true also for the longer
halogen interactions. The phosphate oxygens tend to form straighter
and shorter X-bonds than other electron donors, possibly due to their
negative charge.

In NA$\cdots$ligand noncovalent recognition, a variety of
interactions plays a role. H-bonding and London dispersion-driven
stacking interactions belong to the classic interactions,
complemented by salt bridges, water-network disruptions, or 
metal-ion interactions. Strictly speaking, no noncovalent
ligand has been found to employ X-bonding in the NA binding
(according to the IUPAC recommendations).
No X-bonded NA ligand found is an intriguing fact, which may be
explained in two ways: 

i) The X-bonding ligands may be inefficient in NA recognition.
If so, the halogenated ligands that bind into NAs bind due
to other kinds of interactions that are stronger than X-bonds.
Our data support this hypothesis.
All of the halogen interactions of ligands found in the survey
here show sub-optimal geometries.
Given the high X-bond directionality \cite{Kolar14c},
the stabilization energy of a complex reduces notably when 
deviating from the ideal geometry.
Hence, the contribution to the stabilization of the NA complexes
is likely not dominant. Of course, there are exceptions where 
X-bonding is the driving interaction. For instance, it was proven
that a single X-bond/H-bond exchange might transform the DNA 
conformation completely \cite{Voth07}.

ii) Medicinal chemists may under-appreciate the role of X-bonding in
the drug-design strategies. Our data support also this hypothesis,
because we found only a few halogenated ligands: 20 chlorinated,
8 brominated and one iodinated.

The current study on its own is unable to dissect which reason
is more likely. The starting set of PDBs was biased towards
the lighter halogens, which are, however, less suitable for strong
X-bonds. For a better understanding of the role of X-bonding,
structural and functional studies are required, especially 
of the less-frequent brominated and iodinated compounds.

The effects of halogens are not limited to the X-bonding, though.
The halogenation affects the global electron distribution and
consequently many of the molecular properties.
For instance, Fanfrl\'{i}k \emph{et al.}
demonstrated that solvation/desolvation
may compensate for favorable $\sigma$-hole$\cdots$lone pair 
interaction in halogen-to-hydrogen substitution in a protein-ligand
complex \cite{Fanfrlik15}.
Also, a PDB survey of two protein families revealed shortening on H-bonds
proximal to ligand halogens \cite{Poznanski14}. This corresponds
with the notion that halogenation increases the acidity of proximal
H-bond donors. Still, these effects in NAs are difficult to inspect 
with the limited statistical sample of X-bonds here, and of halogenated
NA ligands in general.

Nevertheless, the current analyses may help in designing
novel X-bonded NA binders. Such
binders should contain a bromine or iodine to form a strong X-bond. 
The ligand binding should be directed to 
the vicinity of sugar-phosphate backbone, for example to groove
or bulge regions. Consequently, such a ligand should likely not
contain extensive aromatic systems that are prone to intercalation
into canonical helices. Systematic halogenation of known
pharmacologically active compounds may be an optimal strategy to
identify new agents with enhanced activity, presumably supported
by the specific X-bonds.

\section{Biographies}

\textbf{Michal H. Kol\'a\v{r}} received his Ph.D. in 2013 from Charles
University in Prague, and from the Institute of Organic Chemistry 
and Biochemistry, Academy of Sciences of the Czech Republic. With 
Pavel Hobza he focused on theoretical and computational description
of noncovalent interactions. He is a recipient of the Humboldt
Research Fellowship for Postdoctoral Researchers. From 2014 he 
worked with Paolo Carloni in Forschungszentrum J\"ulich, Germany,
on RNA-ligand recognition. In 2016 he moved to Helmut Grubm\"uller's
group at the Max Planck Institute for Biophysical 
Chemistry, G\"ottingen, Germany, pursuing his interest in 
non-coding RNAs and computer simulations.

\textbf{Oriana Tabarrini} is a Professor of Medicinal Chemistry
at the Department of Pharmaceutical Sciences (University of Perugia,
Italy). After the degree in Pharmaceutical Chemistry and Technology
she has worked for several years with research grants from 
pharmaceutical industries. In 2002 was promoted to Associate 
Professor. Her research has mainly aimed at developing small 
molecules as pharmacological tools and potential chemotherapeutics
with particular focus on nucleic acid binders as antivirals and more
recently for the treatment of neurodegenerative disorders.
She has published over 85 research articles in leading peer-reviewed
journals, including some invited reviews and patents. She has also
received several project grants and is an ad-hoc reviewer for 
several top journals.
\section{Abbreviations}

\begin{description}
\item[BRU] 5-bromouridine
\item[CLM] chloramphenicol
\item[CLY] clindamycine
\item[CPT] cisplatin
\item[DNA] deoxyribonucleic acid
\item[dsDNA] double-stranded DNA
\item[G4] G-quadruplex
\item[H-bond] hydrogen bond
\item[HCV] Hepatitis C virus
\item[HIV-1] human immunodeficiency virus type 1
\item[IRES] internal ribosome entry site
\item[NA] nucleic acid
\item[ncRNA] non-coding RNA
\item[PDB] Protein Data Bank
\item[PFV] prototype foamy virus
\item[RNA] ribonucleic acid
\item[TAR] trans-activation response RNA element
\item[X-bond] halogen bond
\end{description}

\begin{acknowledgement}

MHK is thankful to N. Leioatts and A. Obarska-Kosinska for valuable 
discussions on the
manuscript, and acknowledges the support provided by the 
Alexander von Humboldt Foundation. Most of the analyses prospered
from Numpy \cite{Vanderwalt11} and Matplotlib \cite{Hunter07} python
libraries.

\end{acknowledgement}




\section{Corresponding Author Information}
Email: michal@mhko.science

\bibliography{refs-mod}

\clearpage

\section{Graphical TOC Entry}
\centering{\includegraphics{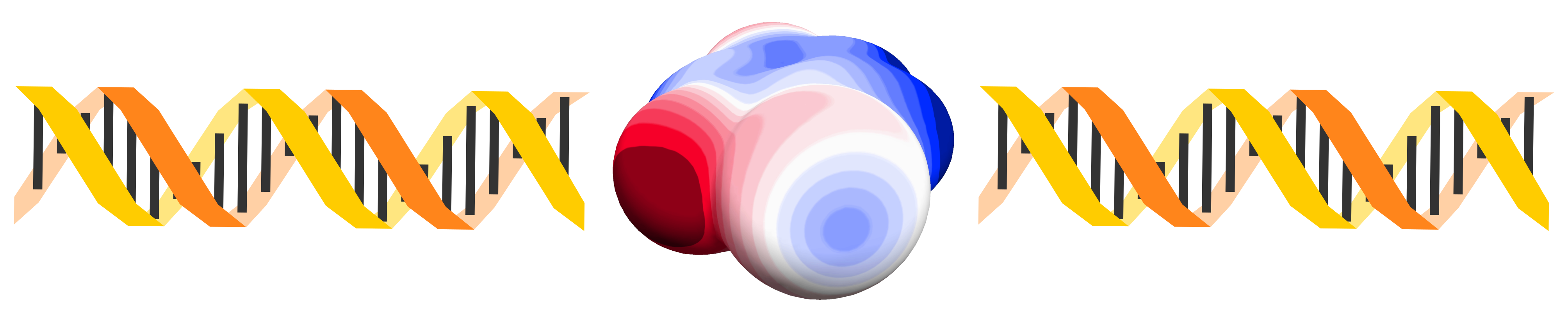}}

\end{document}